\documentclass{aastex}
\usepackage{spr-astr-addons}
\usepackage{url}

\RequirePackage{color}

\usepackage[latin1,applemac]{inputenc}
\usepackage[T1]{fontenc}
\usepackage[english]{babel}
\usepackage{epsfig}
\usepackage{graphicx}
\usepackage{amsmath}
\usepackage{amsfonts}
\usepackage{amsthm}
\usepackage{pstricks}
\usepackage{subfigure}
\usepackage{color}
\usepackage{txfonts}

\begin{document}

\title{Bianchi Type-I Universe Models with Nonlinear Viscosity}

\author{Nouraddin Mostafapoor} \and \author{Øyvind Grøn\altaffilmark{1,2}}

\altaffiltext{1}{Department of Physics,   University of Oslo, N-0316 Oslo, Norway}
\altaffiltext{2}{Oslo College, Faculty of engineering, Pilestredet 35, N-0167 Oslo, Norway}

\maketitle

\begin{abstract}
In this paper we study the evolution of spatially homogeneous and anisotropic Bianchi type-I Universe models with the cosmological constant, $\Lambda$, and filled with nonlinear viscous fluid. The dynamical equations for these models are obtained and solved for some special cases. We calculate the statefinder parameters for the models and display them in the $s$-$r$-plane.
\end{abstract}

\keywords{Bianchi Type-I Universe Model $\cdot$ Bulk and shear viscosity $\cdot$ Scale factor $\cdot$ Hubble parameter $\cdot$ Energy density $\cdot$ Deceleration parameter $\cdot$ Statefinder parameters
}


\section{Introduction}

Astronomical observations indicate that the Universe is undergoing an accelerated expansion (see refs. ~\cite{Riess}, ~\cite{Perlmutter},~\cite{Tegmark}, ~\cite{Abazajian}, ~\cite{Spergel}, ~\cite{Spergel1}, ~\cite{Spergel2} and ~\cite{Bennett}). Based on these observations we know that the Universe is nearly spatially flat. We also know that the Universe consists of about $73\%$ dark energy, which is a fluid with negative pressure, and is responsible for the acceleration of the expansion of the Universe. This unknown dark energy is physically equivalent to the Lorentz invariant vacuum energy (LIVE) and is almost equally distributed in the universe.

The Universe has been successfully described by Einstein's general theory of relativity. Using this theory we can find different models for the Universe. Friedmann-Robertson-Walker (FRW) models describe spatially homogeneous and isotropic universes. But, the FRW models have higher symmetries than the real Universe, and therefore they are probably poor approximations for the very early universe.

The measurements of the cosmic microwave background (CMB) anisotropy (see refs. ~\cite{Spergel}, ~\cite{Spergel1}, ~\cite{Spergel2} and ~\cite{Bennett}) support the existence of anisotropies at the early universe. Therefore, in order to understand the early stage of the evolution of the universe, spatially homogeneous and anisotropic Bianchi type-I (BI) cosmological models are studied.

Bianchi type-I universe models are the simplest models of anisotropic universes that describe a homogeneous and spatially flat space-time and if filled with perfect fluid with the equation of state $p=w\rho$, $ w < 1$, eventually evolve into a FRW universe (see ~\cite{Jacobs}). The isotropy of the present-day universe makes the BI model a prime candidate for studying the possible effects of an anisotropy in the early universe on modern-day data observations.

Bianchi type-I universe models with viscous fluid have been studied by some cosmologists. The influence of viscosity on Bianchi type-I models has been investigated by ~\cite{Belinsk}, and they found that asymptotically for large times, such Bianchi type-I models will approach an isotropic steady-state universe model with a de Sitter space which expands exponentially. For asymptotically early times they found that there exists a Kasner era in which the effects of matter, radiation and viscosity are negligible. Other authors (see refs. ~\cite{Heller}, ~\cite{Woszczyna}, ~\cite{WoszczynaBet}) have also concluded that anisotropic models have in general a vacuum stage near an unavoidable initial singularity in which the energy-momentum tensor has no influence on the cosmic evolution. But,  ~\cite{Gron1990} has found that in a Bianchi type-I universe model filled with viscous Zel'dovich fluid, the bulk viscosity may remove the initial singularity. He also concluded that the viscosity and also LIVE (see ~\cite{Heller}), has an important role in isotropizing the universe. 

 We will, in this paper, study the influence of viscosity on the evolutions of homogeneous and anisotropic Bianchi type-I cosmological models, filled with nonlinear viscous fluid, both with and without a cosmological constant, $\Lambda$. In the end we will apply the statefinder diagnostics to these models to find which models are in agreement with the observational data.  

The present work generalizes a recent analysis of viscous isotropic FRW-universe models (see  ~\cite{nour}) to anisotropic universe models.


\section{Viscous Fluid Cosmology of Bianchi Type-I}

The gravitational field in the spatially homogeneous and anisotropic case is given by a Bianchi type I metric, and it has the form

\begin{equation}
ds^{2}=dt^{2}-R_{i}^{2}(dx^{i})^{2},\label{eq:1}
\end{equation}
where $R_{1} = a(t)$, $R_{2} = b(t)$, $R_{3} = c(t)$ are the directional scale factors. The energy momentum tensor of a linear viscous field has the form
\begin{align}
T^{\mu}_{\nu}=&\left(\rho + p_{\textrm{eff}}\right)u^{\mu}u_{\nu}-\delta^{\mu}_{\nu}p_{\textrm{eff}} \nonumber\\
&+\eta g^{\mu\beta}\left[u_{\nu;\beta}+u_{\beta;\nu}-u_{\nu}u^{\alpha}u_{\beta;\alpha}-u_{\beta}u^{\alpha}u_{\nu;\alpha}\right],\label{eq:2}
\end{align}
where
\begin{equation}
p_{\textrm{eff}}=p-\left(\xi-\frac{2}{3}\eta\right)u^{\nu}_{;\nu},\label{eq:3}
\end{equation}
is the effective pressure, $p$ pressure and $\rho$ is the energy density. The coefficients of bulk and shear viscosity which are denoted by $\xi$ and $\eta$, respectively, are both positively definite, i.e., $\xi > 0$, $\eta > 0$. In a comoving reference frame with diagonal metric tensor, the non-vanishing components of the energy-momentum tensor of a linear viscous fluid are
\begin{align}
T^{0}_{L0}&= \rho_{L}, \label{eq:4}\\  
T^{i}_{Li}&= -p_{\textrm{eff}} + 2\eta H_{i},\label{eq:5}
\end{align} 
where $H_{i}$ are the directional Hubble parameters.
In the same frame the non-vanishing components of the energy-momentum tensor of nonlinear viscous fluid are 
\begin{align}
T^{0}_{N 0}&=\rho_{N}, \label{eq:6}\\
T^{i}_{N i}&=-p_{N}+\alpha \theta^{2}+\beta\theta H_{i}+\lambda(H_{i})^{2}.\label{eq:7}
\end{align}
Here, $\alpha$, $\beta$ and $\lambda$ are constants and ~\cite{Novello} assume that they satisfy the following constraints relation which will be adopted here too,
\begin{equation}
(3\alpha+\beta)9H^{2}+\lambda\sum_{i=1}^{3}H_{i}^{2}=0.\label{eq:8}
\end{equation}   
We will in this paper consider an energy momentum tensor that is given by
\begin{equation}
T=T_{L}+T_{N},\label{eq:9}
\end{equation}
where $T_{L}$ and $T_{N}$ are the energy momentum tensors of fluids with linear and nonlinear viscosity, respectively. In this case, in the same comoving reference frame with diagonal metric tensor, the non-vanishing components of (\ref{eq:9}), are 
\begin{align}
T^{0}_{\quad 0}&=\rho,\label{eq:10}\\
T^{i}_{\quad i}=&-p+2\eta H_{i}+(3\xi-2\eta)H +9\alpha H^{2}+3\beta HH_{i} \nonumber\\
&+\lambda(H_{i})^{2}, \label{eq:11}
\end{align}
where
\begin{align}
\rho&=\rho_{L}+\rho_{N},\nonumber\\
p&=p_{L}+p_{N}.\nonumber
\end{align}
The pressure $p$ is connected to the energy density by means of an equation of state describing a perfect fluid, given by
\begin{equation}
p = w \rho.\label{eq:12}
\end{equation}

We will now give the definitions of some physical parameters. The volume expansion is
\begin{equation}
\theta\equiv u^{\mu}_{;\mu}=3H,\label{eq:13}
\end{equation} 
where we have introduced a generalized Hubble parameter $H$, in analogy with Hubble parameter in a FRW universe, as
\begin{align}
3H&\equiv \frac{\dot{\tau}}{\tau} \label{eq:14}\\
&= \left(\frac{\dot{a}}{a}+\frac{\dot{b}}{b}+\frac{\dot{c}}{c}\right)\\
&=H_{1}+H_{2}+H_{3}, 
\end{align}
where $\tau$ is the volume scale factor of the BI space-time, and is defined as
\begin{equation}
\tau\equiv abc.
\end{equation}
Letting $\Delta H_{i}=H_{i}-H$ the definition of the anisotropy parameter, $A$, ca be written as (see ~\cite{gron1985})
\begin{equation}
A=\frac{1}{3}\sum_{i=1}^{3}\left(\frac{\Delta H_{i}}{H}\right)^{2}=\frac{1}{9}\sum_{i<j}\left(\frac{H_{i}-H_{j}}{H}\right)^{2},  \label{eq:18}
\end{equation}

This parameter tells how anisotropic the space is, and $A=0$ represents an isotropic space. The shear scalar $\sigma$ is defined by
\begin{equation}
\sigma^{2}\equiv \frac{1}{2} \sigma^{\mu \nu}\sigma_{\mu \nu},\label{eq:19}
\end{equation}
where
\begin{align}
\sigma_{\mu\nu}=&\frac{1}{2}\left[u_{\mu;\alpha}\left(\delta^{\alpha}_{\nu}-u^{\alpha}u_{\nu}\right)+u_{\nu;\alpha}\left(\delta^{\alpha}_{\nu}-u^{\alpha}u_{\nu}\right)\right] \nonumber\\
& -\frac{1}{3}\theta\left(g_{\mu\nu}-u_{\mu}u_{\nu}\right). 
\end{align} 
This gives
\begin{align}
2\sigma^{2}&=\sum_{i=1}^{3}H_{i}^{2}-3H^{2} \label{eq:21}\\
&=\frac{1}{3}A\theta^{2}\\
&=\frac{\dot{a}^{2}}{a^{2}}+\frac{\dot{b}^{2}}{b^{2}}+\frac{\dot{c}^{2}}{c^{2}}-\frac{1}{3}\theta^{2}.
\end{align}

From equations (\ref{eq:8}) and (\ref{eq:18}) follow that 
\begin{equation}
	3(3\alpha+\beta)+ \lambda(1+A) = 0. \label{eq:24}
\end{equation} 
If $\alpha$, $\beta$, $\lambda$ are constant this relationship requires that $A$ is constant, which is not the case in general for Bianchi type I universe models with viscous fluids. Hence from now on we will assume that $\lambda = 0$, and thus also that $\beta = -3\alpha$. Hence if $\alpha>0$ it follows that $\beta<0$.

\subsection{Field equations and their solutions}

The Einstein's field equation with the cosmological constant, $\Lambda$, are given by
\begin{equation}
R^{\mu}_{\nu}-\frac{1}{2}\delta^{\mu}_{\nu}R=T^{\mu}_{\nu}+\delta^{\mu}_{\nu} \Lambda,\label{eq:25}
\end{equation}
where we have set $\kappa=8\pi G=1$. With $\beta = -3\alpha$ and $\lambda = 0$, the field equations take the form

\begin{align}
\frac{\ddot{b}}{b}+\frac{\ddot{c}}{c}+\frac{\dot{b}}{b}\frac{\dot{c}}{c}=&-p+3\xi H+2\eta\Delta H_{1}-9\alpha H \Delta H_{1} +\Lambda, \label{eq:26} 
\end{align}

\begin{align}
\frac{\ddot{a}}{a}+\frac{\ddot{c}}{c}+\frac{\dot{a}}{a}\frac{\dot{c}}{c}=&-p+3\xi H+2\eta\Delta H_{2}-9\alpha H \Delta H_{2} +\Lambda,  \label{eq:27} 
\end{align}

\begin{align}
\frac{\ddot{a}}{a}+\frac{\ddot{b}}{b}+\frac{\dot{a}}{a}\frac{\dot{b}}{b}=&-p+3\xi H+2\eta\Delta H_{3}-9\alpha H \Delta H_{3} +\Lambda, \label{eq:28} 
\end{align}

\begin{equation}
\frac{\dot{a}}{a}\frac{\dot{b}}{b}+\frac{\dot{b}}{b}\frac{\dot{c}}{c}+\frac{\dot{c}}{c}\frac{\dot{a}}{a}=\rho+\Lambda.\label{eq:29}
\end{equation}

Using the definition of the anisotropy parameter, $A$, in equation (\ref{eq:18}), we can write equation (\ref{eq:29}) as
\begin{equation}
\rho=\left(1-\frac{A}{2}\right)3H^{2}-\Lambda.\label{eq:30}
\end{equation}
Expressing this equation by the volume expansion, i.e. $\theta=3H$, and the shear scalar $\sigma$, equation (\ref{eq:19}), we obtain
\begin{equation}
\rho=\frac{1}{3}\theta^{2}-\sigma^{2}-\Lambda.\label{eq:31}
\end{equation}
Solving equations (\ref{eq:26})-(\ref{eq:28}), we find the expressions for the metric functions. By subtracting (\ref{eq:26}) from (\ref{eq:27}), (\ref{eq:26}) from (\ref{eq:28}) and (\ref{eq:27}) from (\ref{eq:28}), one finds

\begin{align}
\frac{\ddot{a}}{a}-\frac{\ddot{b}}{b}+\frac{\dot{c}}{c}\left[\frac{\dot{a}}{a}-\frac{\dot{b}}{b}\right]&=\left[-2\eta+9\alpha H\right]\left[\frac{\dot{a}}{a}-\frac{\dot{b}}{b}\right],\label{eq:32}\\
\frac{\ddot{a}}{a}-\frac{\ddot{c}}{c}+\frac{\dot{b}}{b}\left[\frac{\dot{a}}{a}-\frac{\dot{c}}{c}\right]&=\left[-2\eta+9\alpha H\right]\left[\frac{\dot{a}}{a}-\frac{\dot{c}}{c}\right],\label{eq:33}\\
\frac{\ddot{b}}{b}-\frac{\ddot{c}}{c}+\frac{\dot{a}}{a}\left[\frac{\dot{b}}{b}-\frac{\dot{c}}{c}\right]&=\left[-2\eta+9\alpha H\right]\left[\frac{\dot{b}}{b}-\frac{\dot{c}}{c}\right].\label{eq:34}
\end{align}
We divide (\ref{eq:32}) by $\left[\frac{\dot{a}}{a}-\frac{\dot{b}}{b}\right]$, (\ref{eq:33}) by $\left[\frac{\dot{a}}{a}-\frac{\dot{c}}{c}\right]$ and (\ref{eq:34}) by $\left[\frac{\dot{b}}{b}-\frac{\dot{c}}{c}\right]$, and we obtain
\begin{align}
\frac{\ddot{a}b-\ddot{b}a}{a\dot{b}-\dot{a}b}&=-2\eta+9\alpha H-\frac{\dot{c}}{c}, \label{eq:35}\\
\frac{\ddot{a}c-\ddot{c}a}{a\dot{c}-\dot{a}c}&=-2\eta+9\alpha H-\frac{\dot{b}}{b},\label{eq:36}\\
\frac{\ddot{b}c-\ddot{c}b}{b\dot{c}-\dot{b}c}&=-2\eta+9\alpha H-\frac{\dot{a}}{a}.\label{eq:37}
\end{align}
Integrating equations (\ref{eq:35})-(\ref{eq:37}), we get the following relations between $H_{1}$ and $H_{2}$, $H_{2}$ and $H_{3}$, and $H_{1}$ and $H{3}$
\begin{align}
\frac{\dot{a}}{a}-\frac{\dot{b}}{b}&=X_{1}\tau^{3\alpha-1}\textrm{exp}\left(-\Phi\right),\label{eq:38} \\
\frac{\dot{b}}{b}-\frac{\dot{c}}{c}&=X_{2}\tau^{3\alpha-1}\textrm{exp}\left(-\Phi\right),\label{eq:39} \\
\frac{\dot{a}}{a}-\frac{\dot{c}}{c}&=X_{3}\tau^{3\alpha-1}\textrm{exp}\left(-\Phi\right).\label{eq:40}
\end{align}
Here $X_{1}$, $X_{2}$, $X_{3}$ are integration constants, which satisfy 
\begin{displaymath}
X_{1}+X_{2}-X_{3}=0,
\end{displaymath}
and we have defined 
\begin{displaymath}
\Phi\equiv2\int{\eta}dt.
\end{displaymath}
We have here used
\begin{equation}
\dot{(\textrm{ln}\tau)}=3H, \quad \Rightarrow \tau=\textrm{e}^{3\int_{0}^{t} H\textrm{dt}}, \label{eq:41}
\end{equation}
where we have assumed that $\tau(0)=0$. Integrating equations (\ref{eq:38})-(\ref{eq:40}), we find
\begin{align}
\frac{{a}}{b}&=D_{1}\textrm{exp}\left[X_{1}\int\tau^{3\alpha-1}\textrm{e}^{-\Phi}\textrm{dt}\right],\label{eq:42} \\
\frac{{b}}{c}&=D_{2}\textrm{exp}\left[X_{2}\int\tau^{3\alpha-1}\textrm{e}^{-\Phi}\textrm{dt}\right],\label{eq:43} \\
\frac{a}{c}&=D_{3}\textrm{exp}\left[X_{3}\int\tau^{3\alpha-1}\textrm{e}^{-\Phi}\textrm{dt}\right],\label{eq:44}
\end{align}
where $D_{1}$, $D_{2}$, $D_{3}$ are integration constants.
From the equations (\ref{eq:42})-(\ref{eq:44}) we can write the metric functions as
\begin{align}
a(t)&=A_{1}\tau^{1/3}\textrm{exp}\left[\int\tau^{3\alpha-1}\textrm{e}^{-\Phi}\left(\frac{X_{1}}{3}+\frac{X_{3}}{3}\right)\textrm{dt}\right],\label{eq:45} \\
b(t)&=A_{2}\tau^{1/3}\textrm{exp}\left[\int\tau^{3\alpha-1}\textrm{e}^{-\Phi}\left(\frac{X_{2}}{3}-\frac{X_{1}}{3}\right)\textrm{dt}\right],\label{eq:46} \\
c(t)&=A_{3}\tau^{1/3}\textrm{exp}\left[-\int\tau^{3\alpha-1}\textrm{e}^{-\Phi}\left(\frac{X_{3}}{3}+\frac{X_{2}}{3}\right)\textrm{dt}\right],\label{eq:47}
\end{align}
where
\begin{displaymath}
A_{1} = \sqrt[3]{(D_{1}D_{3})}, \quad A_{2} = \sqrt[3]{(D_{2}/D_{1})}, \quad A_{3}=\sqrt[3]{1/(D_{2}D_{3})}.
\end{displaymath}

We will now go back to equations (\ref{eq:38})-(\ref{eq:40}) and rewrite them in compact form as
\begin{equation}
H_{i} - H_{j} = X_{k}\tau^{3\alpha-1} \text{e}^{-\Phi}. \label{eq:48}
\end{equation}

Using equation (\ref{eq:8}) and adding Einstein equations (\ref{eq:26}), (\ref{eq:27}) and (\ref{eq:28}), we obtain the Raychaudhury equation 
\begin{equation}
\dot{H}=-3H^{2}+\frac{1}{2}(1-w)\rho+\frac{3}{2}\xi H+\Lambda, \label{eq:49}
\end{equation}
where we have used the equation of state
\begin{displaymath}
p=w\rho. 
\end{displaymath}

The deSitter spacetime has $\rho$=$\xi$=$\dot{H}$=0 and $H=\sqrt{\frac{\Lambda}{3}}$ with $\Lambda >0$.

\subsection{Energy Conservation Equation}

In a comoving reference frame with diagonal metric tensor the equation of energy conservation $T^{\nu}_{0;\nu}=0$ may be written as
\begin{equation}
\dot{T^{0}_{0}}+\dot{\left(\textrm{ln}\sqrt{-g}\right)}T^{0}_{0}-\frac{1}{2}\dot{g_{\alpha\alpha}}T^{\alpha\alpha}=0.
\end{equation}
Inserting the components of the energy momentum tensor in equation (\ref{eq:11}), we obtain
\begin{align}
\dot{\rho}+3H(\rho+p)=&3(3\xi-2\eta)H^{2}+2\eta\sum_{i=1}^{3}H_{i}^{2}+27\alpha H^{3} \nonumber\\
&-9\alpha H\sum_{i=1}^{3}H_{i}^{2}.
\end{align}
Using the definition of the anisotropy parameter, we can rewrite this equation as 
\begin{align}
\dot{\rho}+3H(\rho+p)=&3(3\xi+2\eta A)H^{2}-27\alpha AH^{3}.\label{eq:52}
\end{align}
From this equation we can see that at the early stage of the cosmic expansion, when the Hubble parameter has a large value, viscosity has an important role in energy production. If $p=-\rho$ then $\dot{\rho}>0$.


\section{Statefinder Formalism for Bianchi Type-I Universe Models with Nonlinear Viscous Fluid}

The statefinder parameter pair $\left\{s,r\right\}$, was introduced by ~\cite{Sahni} and ~\cite{Alam}, and for a flat FRW universe they are defined as

\begin{align}
r&\equiv \frac{\dddot{a}}{aH^{3}},\label{eq:53}\\
s&\equiv \frac{r-1}{3(q-1/2)}.\label{eq:54}
\end{align}

The statefinder diagnostic has a geometrical character, because it is constructed from the space-time metric directly, which is more universal than "physical" variables, which are model-dependent. Introducing the statefinder parameters is a natural next step beyond the Hubble parameter $H$ depending on $\dot{a}$ and the deceleration parameter $q$ depending on $\ddot{a}$. For a FRW-universe the definition of the deceleration parameter is 
\begin{equation}
q = -\frac{\ddot{a}}{aH^{2}}.\label{eq:55}
\end{equation}
Expressing the deceleration parameter and the statefinder parameters in terms of the Hubble parameter and its derivatives with respect to cosmic time, we obtain (see reference ~\cite{Gron2005})
\begin{align}
q&=-1-\frac{\dot{H}}{H^{2}},\label{eq:56}\\
r&=1+3\frac{\dot{H}}{H^{2}}+\frac{\ddot{H}}{H^{3}},\label{eq:57}\\ 
s&=-\frac{2}{3H}\frac{3H\dot{H}+\ddot{H}}{3H^{2}+2\dot{H}}.\label{eq:58}
\end{align}

The corresponding parameters for a Bianchi type I universe are defined by equations (\ref{eq:56})-(\ref{eq:58}), where the generalized Hubble parameter H is defined in equation (\ref{eq:13}). In order to calculate these parameters we first differentiate equation (\ref{eq:49}), and obtain
\begin{equation}
\ddot{H}=-6H\dot{H}+\frac{1}{2}(\dot{\rho}-\dot{p})+\frac{3}{2}\left(\dot{\xi} H+\xi\dot{H}\right).\label{eq:59}
\end{equation} 
We find the general expressions for the deceleration parameter and statefinder parameters for Bianchi type-I universe models with nonlinear viscous fluid by inserting equation (\ref{eq:49}) and equation (\ref{eq:59}) in equations (\ref{eq:56})-(\ref{eq:57}), and obtain 
\begin{equation}
q=2-\frac{3}{2}\frac{\xi}{H}-\frac{1}{2}\frac{(1-w)\rho}{H^{2}}-\frac{\Lambda}{H^{2}}, \label{eq:60}
\end{equation}
\begin{align}
r=&10-9\frac{\xi}{H}+\frac{1}{2}(\dot{\rho}-\dot{p})\frac{1}{H^{3}}+\frac{3}{2}(\rho-p)(\frac{1}{2}\xi-H)\frac{1}{H^{3}}\nonumber\\
&+\frac{3}{2}\frac{\dot{\xi}-\frac{3}{2}\xi^{2}}{H^{2}}+\frac{\Lambda}{H^{2}}\left(1-\frac{3}{2}\frac{\xi}{H}\right),\label{eq:61}
\end{align}

From these expressions, we can see that if we know the equation of state, the deceleration parameter and statefinder parameter are related to the quantities $H$, $\rho$, $\xi$, $\dot{\xi}$ and $\Lambda$. 

Using equation (\ref{eq:60}) and defining the density parameters of the dark matter and dark energy represented by the cosmological constant, $\Lambda$, with present values, as
\begin{equation}
\Omega_{m0}=\frac{ \rho_{m0}}{3H_{0}^{2}}, \qquad \Omega_{\Lambda 0}=\frac{\Lambda}{3H_{0}^{2}},
\end{equation}
and introducing a dimensionless $viscosity$ $parameter$, $\Omega_{\xi}$, with present value
\begin{equation}
\Omega_{\xi 0}=\frac{3 \xi}{H_{0}},
\end{equation}
we obtain
\begin{equation}
q = 2 -\frac{3}{2}(1-w)\Omega_{m}-\frac{1}{2}\Omega_{\xi}-3\Omega_{\Lambda }.
\end{equation}
In terms of anisotropic parameter this expression takes the form
 \begin{equation}
2q = 4-3(1-w)(1-\frac{A}{2})-3(1+w)\Omega_{\Lambda}-\Omega_{\xi}.
\end{equation}
We rewrite this equation to obtain
 \begin{equation}
\Omega_{\xi} = 4-3(1-w)(1-\frac{A}{2})-3(1+w)\Omega_{\Lambda}-2q.\label{eq:66}
\end{equation}
Equation (\ref{eq:66}) generalizes our previous work on viscous isotropic FRW-universe models (see ~\cite{nour}) to anisotropic universe models. From this equation we see that by measuring the deceleration parameter and the content of the dark energy, $\Omega_{\Lambda}$, in the universe we may obtain information about the amount of viscosity in the cosmic fluid as was shown by ~\cite{pavon2}. Assuming that the universe is isotropic at the present time and $w=0$ for pressure-less matter, we get
\begin{equation}
\Omega_{\xi 0}=\Omega_{m0}-2\Omega_{\Lambda 0}-2q_{0}.
\end{equation}
If there is no mechanism producing viscosity, $\Omega_{\xi 0}=0$ and $q_{0}=(1/2)(\Omega_{m0}-2\Omega_{\Lambda 0})$. As it was mentioned in ~\cite{nour} at the present time we have very accurate measurements of $\Omega_{m0}$ and $\Omega_{\Lambda 0}$, but there is a relatively great uncertainty in the kinematical measurements of the time variation of the Hubble parameter, i.e., of $q_{0}$. So at the present time we have no accurate information from such measurements about the importance of cosmic viscosity. ~\cite{zim3} have shown, however, that cosmic particle production can produce an effective bulk viscosity which may be of significance for the explanation of the accelerated expansion of the universe.


\section{Anisotropy Parameter}

Differentiating equation (\ref{eq:30}), we obtain
\begin{equation}	
	\dot{\rho} = -\frac{3}{2}H^{2}\dot{A} + (1 - \frac{A}{2})6H\dot{H}. \label{eq:68}
\end{equation}
Inserting equations (\ref{eq:49}) and (\ref{eq:52}) in equation (\ref{eq:68}), we obtain
\begin{align}
	\frac{d}{dt}(\ln(A)) = &\left[\frac{1}{2}(w-1)(2-A)+6\alpha \right] 3H \nonumber\\
	&-4\eta -3\xi -\frac{(1+w)\Lambda}{H}.\label{eq:69}
\end{align}


\section{General Solutions for Bianchi Type I Universe Models with Nonlinear Viscosity}

Inserting equation (\ref{eq:48}) into equation (\ref{eq:18}) we obtain
\begin{equation}
	A = C \frac{\tau^{2(3\alpha-1)}\text{e}^{-2\Phi}}{9H^{2}}, \label{eq:70}
\end{equation}
where $C \equiv X_{1}^{2} + X_{2}^{2} + X_{3}^{2}$. In order to be able to take the limit of a linear viscous fluid we shall assume that $\alpha <1/3$. With the expression (\ref{eq:70}) for the anisotropy parameter equation (\ref{eq:30}) takes the form
\begin{equation}
	\rho = 3 H^{2} - \frac{C}{6} \tau^{2(3\alpha-1)}\text{e}^{-2\Phi} -\Lambda. \label{eq:71}
\end{equation}
From equations (\ref{eq:70}) and (\ref{eq:71}) we see that the shear viscosity contributes to not only the isotropization but also the energy production of the universe.


\section{Solutions for a Bianchi Type I Universe Filled with Zel'dovich Fluid in the Presence of a Cosmological Constant}

For a Zel'dovich fluid with $w = 1$ equation (\ref{eq:69}) reduces to 
\begin{equation}
	\frac{d}{dt}(\ln(A)) = 18\alpha H -4\eta -3\xi - 2 \frac{\Lambda}{H}.\label{eq:72}
\end{equation}
Assuming $\Lambda=0$ and that $\beta$, $\eta$ and $\xi$ are constants, we can integrate equation (\ref{eq:72}) from the point of time $t=0$ of the Big Bang to an arbitrary time $t$, obtaining
\begin{equation}
	A (t) = A_{0}\tau^{6\alpha} \text{e}^{-(4\eta + 3\xi) t}.\label{eq:73}
\end{equation}
We see that the shear and bulk viscosities contribute to isotropization of the universe. With $\Lambda=0$ and no viscosity ~\cite{gron1985} has shown that the anisotropy parameter has constant value $A=2$. We therefore choose $A_{0}=2$.

With $w=1$ equation (\ref{eq:49}) reduces to
\begin{equation}
	\dot{H} = -3H^{2} + \frac{3}{2}\xi H+\Lambda.  \label{eq:74}
\end{equation}
Assuming that the bulk viscosity is constant, we can solve this equation, and we get the following expression for the Hubble parameter
\begin{equation}
	H(t)=\frac{\xi}{4} + \hat{H}\coth{\left[3\hat{H}t\right]}, \label{eq:75}
\end{equation} 
where
\begin{displaymath}
	\hat{H}^{2} \equiv\left(\frac{\xi}{4}\right)^{2}+H_{\Lambda}^{2} \quad \text{and} \quad  H_{\Lambda}^{2}\equiv\frac{\Lambda}{3}.
\end{displaymath}
The volume scale factor  of the BI space-time, normalized to unity at the present time, $t_{0}$, now takes the form
\begin{equation}
	\tau(t)=e^{\frac{3\xi}{4}(t-t_{0})}\frac{\sinh{\left[3\hat{H}t\right]}}{\sinh{\left[3\hat{H}t_{0}\right]}}. \label{eq:76}
\end{equation} 
Now that we know the expressions for the $H$ and $\tau$, using equations (\ref{eq:70}) and (\ref{eq:71}) we can obtain expressions for the anisotropy parameter and the energy density of the Zel'dovich fluid. In the case of a linear viscous fluid the anisotropy parameter is 
\begin{equation}
	A = 2 e^{-(4\eta+3\xi)t}.\label{eq:77}
\end{equation}

\begin{figure}[h!]
  \centering
  \subfigure[][$H(t)$.]{\label{fig:fig1a}\includegraphics[width=0.45\textwidth]{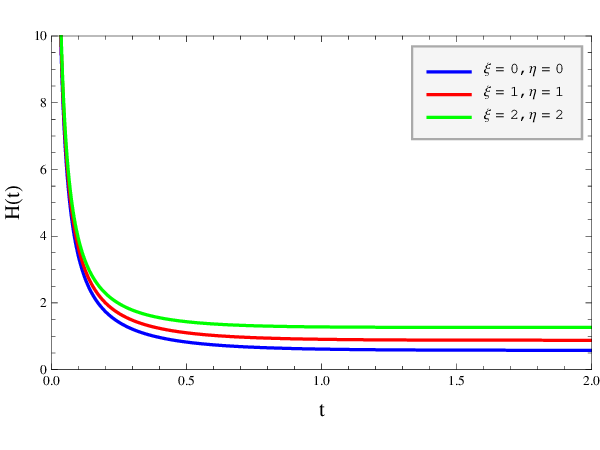}}
  \subfigure[][$\tau(t)$.]{\label{fig:fig1b}\includegraphics[width=0.45\textwidth]{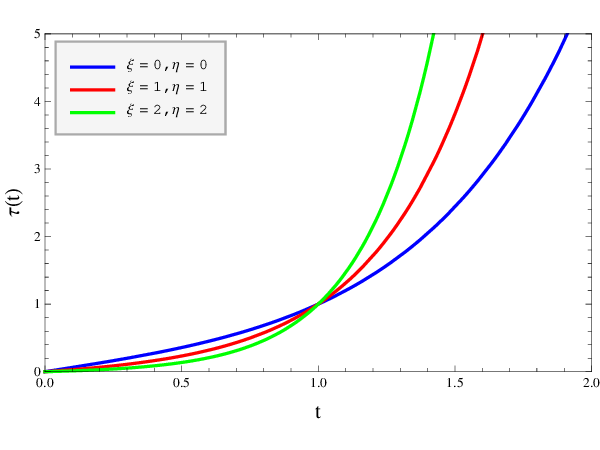}}
   \subfigure[][$A(t)$.]{\label{fig:fig1c}\includegraphics[width=0.45\textwidth]{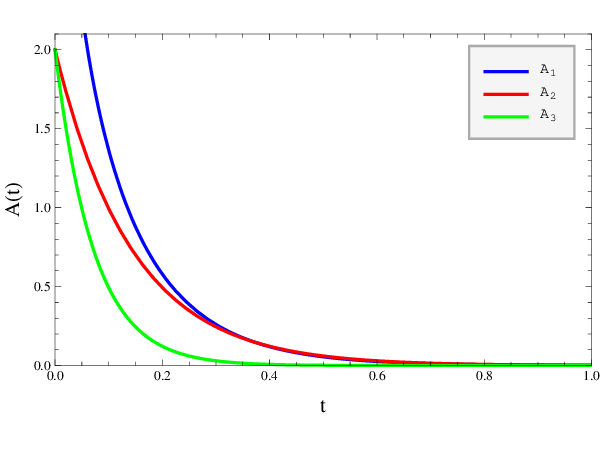}}
  \caption[The Hubble parameter]{The Hubble parameter, the volume scale factor and the anisotropy parameter as functions of time for Bianchi type 1 model plotted in Fig.\ref{fig:fig1a}, \ref{fig:fig1b} and \ref{fig:fig1c}, respectively. Here $w=1$ and $\Lambda=1.0$ in all cases, and $A_{1} $: $\beta=0.1$, $\xi=1.0$, $\eta=1.0$, $A_{2} $: $\beta=0.0$, $\xi=1.0$, $\eta=1.0$ and $A_{3} $: $\beta=0.0$, $\xi=2.0$, $\eta=2.0$.}
  \label{fig:fig1}
\end{figure}

\begin{figure}[h!]
  \centering
  \includegraphics[width=0.45\textwidth]{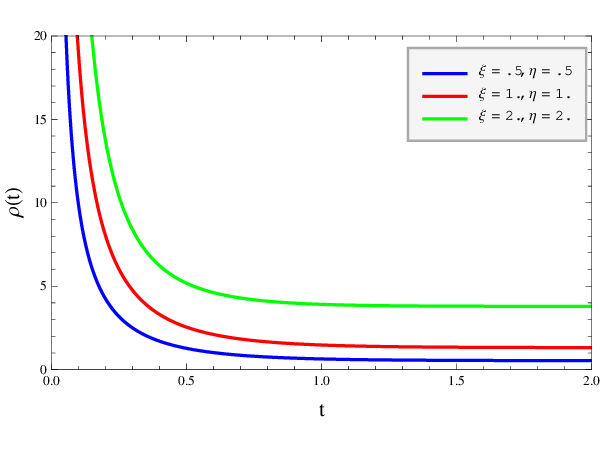}
  \caption[The energy density]{The energy density as a function of time. Here $w=1$, $\Lambda=1.0$ and $\beta=0.0$.}
  \label{fig:fig2}
\end{figure}

In figure \ref{fig:fig1} and figure \ref{fig:fig2} we have plotted the Hubble parameter, the energy density, the volume scale factor and the anisotropy parameter as functions of time for some appropriate values of $\Lambda$, $\xi$, $\beta$ and $\eta$. These figures show that both the Hubble parameter and the energy density are infinitely large at the beginning of the cosmic evolution. As $t$ increases the Hubble parameter and the energy density will decrease and will, eventually, approach finite values. The volume scale factor is zero at $t=0$, and as $t$ increases it will also increase. For $\beta = 0$ the anisotropy parameter has the value $2$ at $t=0$, and,  as $t$ increases it will tend to zero as $t\rightarrow \infty$. The bigger the values of shear and bulk viscosities are the faster the anisotropy parameter goes to zero. This means that the shear and bulk viscosities contribute to isotropization of the Universe.
   
   From the evolution of the energy density we see that if the values of $\xi$ and $\eta$ increase, the energy density will also increase. Therefore, it follows that the presence of nonlinear fluid and the shear and the bulk viscosities contribute to energy production of the universe. 
         
\begin{figure}[h!]
  \centering
  \subfigure[][$q(t)$. Here $w=1$, $\beta=0$ and $\Lambda=1$.]{\label{fig:fig3a}\includegraphics[width=0.45\textwidth]{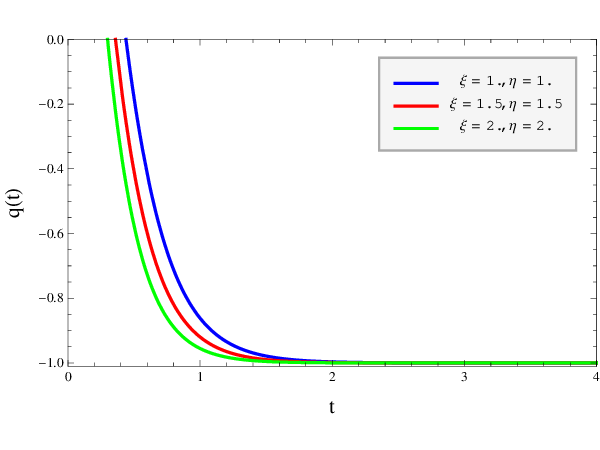}}
  \subfigure[][$q(t)$. Here $w=1$, $\beta=0$, $\xi=1$ and $\eta=1$.]{\label{fig:fig3b}\includegraphics[width=0.45\textwidth]{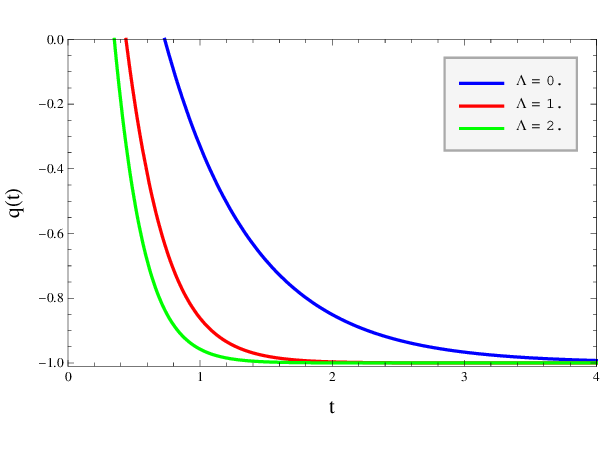}}
  \caption[The deceleration parameter.]{The deceleration parameter as a function of time.}
  \label{fig:fig3}
\end{figure}

\begin{figure}[h!]
  \centering
  \subfigure[][The r-s-plane. Here $w=1$, $\beta=0$, $\xi=1$ and $\eta=1$.]{\label{fig:fig4a}\includegraphics[width=0.45\textwidth]{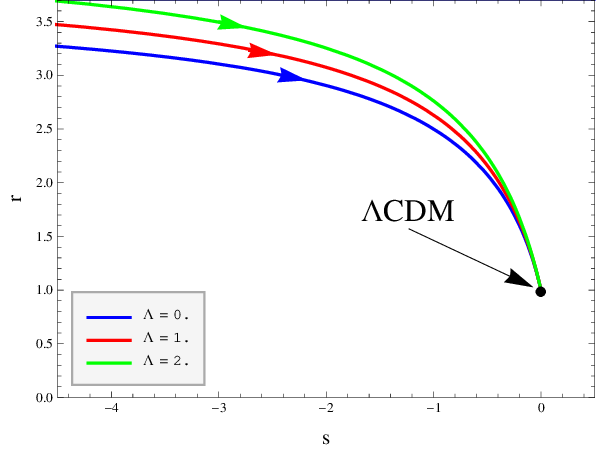}}
  \subfigure[][The r-s-plane. Here $w=1$, $\beta=0$ and $\Lambda=1$.]{\label{fig:fig4b}\includegraphics[width=0.45\textwidth]{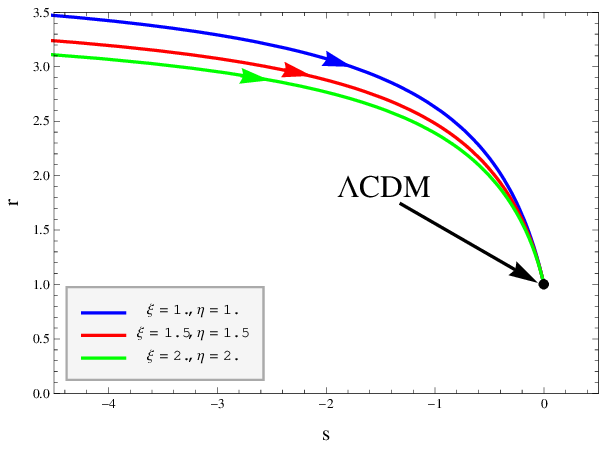}}
  \caption[r-s-plane]{ The $r-s$-plane. The point $\{0,1\}$ corresponds to $\{s,r\}$ for the $\Lambda\textrm{CDM}$ model with no viscosity. }
  \label{fig:fig4}
\end{figure}

   The evolution of the deceleration parameter and the $r$-$s$-plane for this solution are displayed in figure \ref{fig:fig3} and figure \ref{fig:fig4}. From these figures we see that the statefinder parameters will approach $\left\{s,r\right\}=\left\{0,1\right\}$ as $t\rightarrow t_{0}$, which is the values for the $\Lambda \textrm{CDM}$ universe model. It means that these models will eventually become more like the $\Lambda \textrm{CDM}$ universe model as $t\rightarrow t_{0}$. 

\section{Solutions for the Bianchi Type I Universe in the Limit $A\rightarrow 0$}

Now, inserting equation (\ref{eq:30}) into equation (\ref{eq:49}), we obtain
\begin{equation}
	\dot{H} = \left[ \frac{1}{2}(1-w)(1-\frac{A}{2}) - 1 \right] 3H^{2} + \frac{3}{2}\xi H + \frac{(1+w)\Lambda}{2}. \label{eq:78} 
\end{equation}
Assuming that $A\rightarrow 0$ at late times, equation (\ref{eq:78}) reduces to 
\begin{equation}
	\dot{H} = -\frac{3}{2}(1+w)H^{2} + \frac{3}{2}\xi H + \frac{(1+w)\Lambda}{2}. \label{eq:79} 
\end{equation}

In the limit $A\rightarrow 0$ equation (\ref{eq:69}) reduces to 
\begin{equation}
	(1+w)\Lambda = \left[(6\alpha+w-1)3H-4\eta-3\xi\right]H. \label{eq:80}
\end{equation}
Inserting this into equation (\ref{eq:79}) we obtain
\begin{equation}
\dot{H} = 3 (3\alpha-1)H^{2}-2\eta H. \label{eq:81}
\end{equation}
Integrating equation (\ref{eq:81}) we get 
\begin{equation}
H(t) = \frac{H_{0}}{\left[1+ \frac{3}{2}\frac{1-3\alpha}{\eta}H_{0}  \right]\text{e}^{2\eta(t-t_{0})}-\frac{3}{2}\frac{1-3\alpha}{\eta}H_{0}}. \label{eq:82}
\end{equation}
where $H_{0} = H (t_{0})$. A new integration with the average scale factor $a = \tau^{1/3}$, normalized to unity at the present time, $t_{0}$, and assuming that $a(0)=0$, leads to 
\begin{equation}
	\text{e}^{2\eta t_{0}} =1+ \frac{2\eta}{3(1-3\alpha)H_{0}} . \label{eq:83}
\end{equation}
and
\begin{equation}
a(t) = \left[\frac{1-\text{e}^{-2\eta t}}{1-\text{e}^{-2\eta t_{0}}}\right]^{\frac{1}{3(1-3\alpha)}}. \label{eq:84}
\end{equation}
The Hubble parameter in this case takes the form
\begin{equation}
	H(t)=\frac{\text{e}^{2\eta t_{0}}-1}{\text{e}^{2\eta t}-1}H_{0}. \label{eq:85}
\end{equation}

\begin{figure}[h!]
  \centering
  \subfigure[][$H(t)$. Here $\eta=1$.]{\label{fig:fig5a}\includegraphics[width=0.45\textwidth]{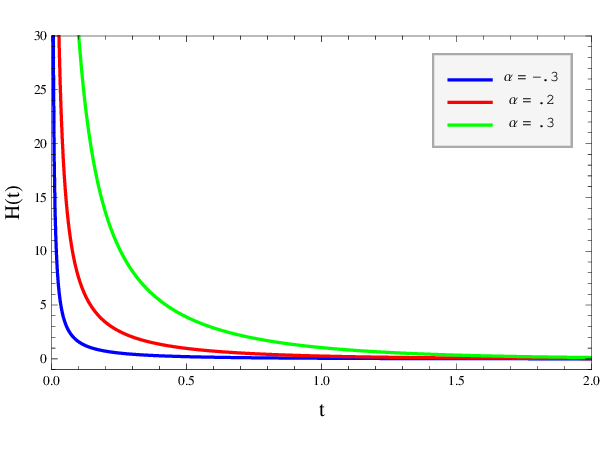}}
  \subfigure[][$H(t)$. Here $\alpha=0.3$.]{\label{fig:fig5b}\includegraphics[width=0.45\textwidth]{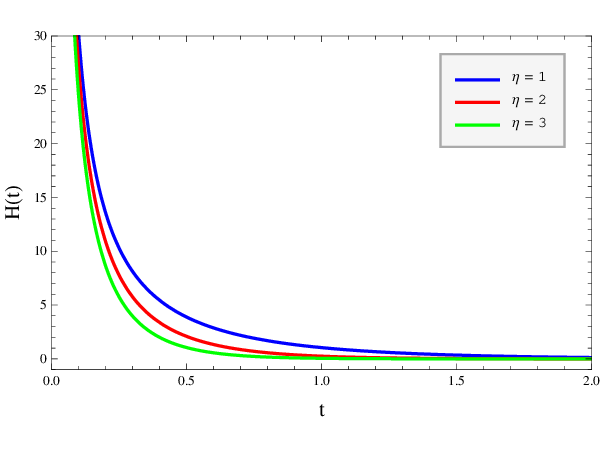}}
  \caption[Hubble]{The Hubble parameter as a function of time.}
  \label{fig:fig5}
\end{figure}

\begin{figure}[h!]
  \centering
  \subfigure[][$a(t)$. Here $\eta=1$.]{\label{fig:fig6a}\includegraphics[width=0.45\textwidth]{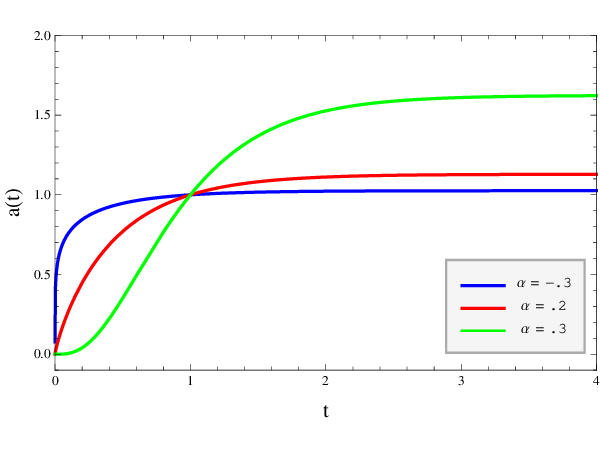}}
  \subfigure[][$a(t)$. Here $\alpha=0.3$.]{\label{fig:fig6b}\includegraphics[width=0.45\textwidth]{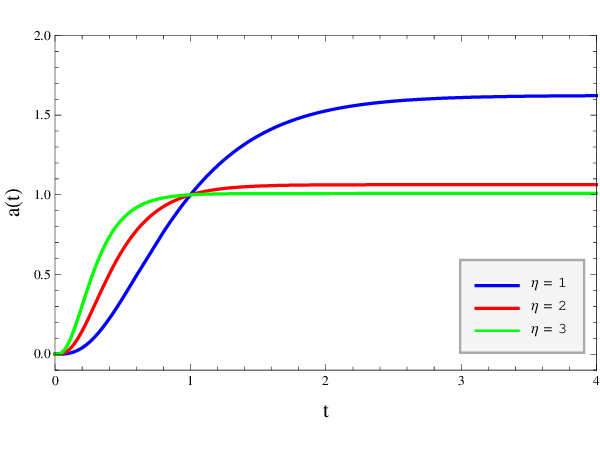}}
  \caption[scale factor]{The average scale factor as a function of time.}
  \label{fig:fig6}
\end{figure}

\begin{figure}[h!]
  \centering
  \subfigure[][$\rho(t)$. Here $\eta=1$.]{\label{fig:fig7a}\includegraphics[width=0.45\textwidth]{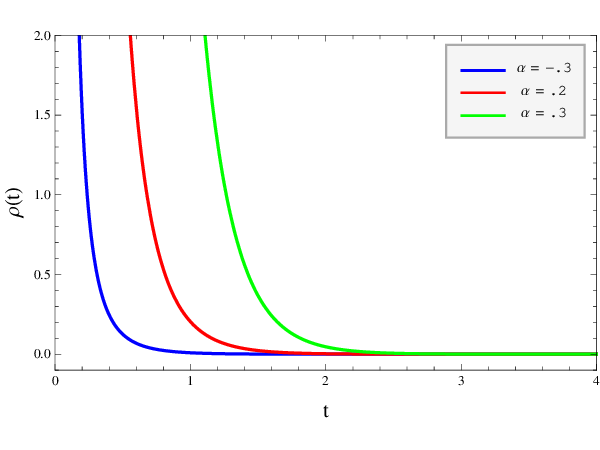}}
  \subfigure[][$\rho(t)$. Here $\alpha=0.3$.]{\label{fig:fig7b}\includegraphics[width=0.45\textwidth]{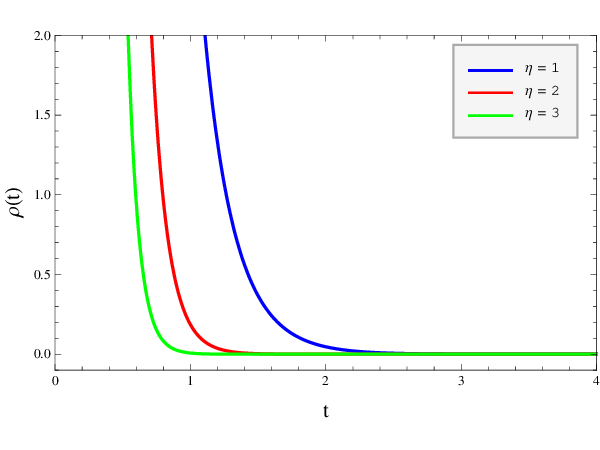}}
  \caption[energy density]{The energy density as a function of time.}
  \label{fig:fig7}
\end{figure}

In the limit $t\rightarrow \infty$, we see from equation (\ref{eq:84}) that $a(t)$ approaches 
\begin{equation}
	a_{\text{max}} =  \left[1-\text{e}^{-2\eta t_{0}}\right]^{\frac{1}{3(3\alpha-1)}}. \label{eq:86}
\end{equation}
In figures \ref{fig:fig5}, \ref{fig:fig6} and \ref{fig:fig7} we have plotted the Hubble parameter, the energy density and the average scale factor as functions of time for different values of $\alpha$ and $\eta$, respectively. From these figures we see that both the Hubble parameter and the energy density tend to zero as $t\rightarrow\infty$.  The volume of the universe will increase as $t\rightarrow t_{0}$,  but, as $t\rightarrow \infty$ it will approach a finite value. From these results we can conclude that as $t\rightarrow\infty$ this universe will be empty and it will stop expanding.

\section{Solutions for the Case with Variable Shear Viscosity}

Inserting equation (\ref{eq:71}) into (\ref{eq:49}), we obtain
\begin{equation}
\dot{H} = -\frac{3}{2}(1+w)H^{2}+\frac{3}{2}\xi H- \frac{C}{12}(1-w)\tau^{2(3\alpha-1)}\text{e}^{-2\Phi}+\frac{1}{2}(1+w)\Lambda. \label{eq:87}
\end{equation}
We will first consider the case with shear viscosity being proportional to the Hubble parameter, given by
\begin{equation}
\eta = -\frac{3}{2}(1-3\alpha)H.\label{eq:88}
\end{equation}
In order to solve Raychaudhury equation we have to specify the form of the bulk viscosity, $\xi$. Some authors (see ~\cite{WBXHU} and ~\cite{MGHUXHM}) have proposed a possible form of bulk viscosity as
\begin{equation}
\xi=\xi_{0}+\xi_{1}\frac{\dot{\tau}}{\tau}+\xi_{2}\frac{\ddot{\tau}}{\dot{\tau}}. \label{eq:89}
\end{equation}
Using that 
\begin{displaymath}
3\dot{H}=\frac{d}{dt}\left(\frac{\dot{\tau}}{\tau}\right)=\frac{\ddot{\tau}}{\tau}-9H^{2},
\end{displaymath}
the bulk viscosity in equation (\ref{eq:89}) takes the form
\begin{equation}
\xi=\xi_{0}+3\xi_{1}H+\xi_{2}\left(\frac{\dot{H}}{H}+3H\right). \label{eq:90}
\end{equation}
Inserting equations (\ref{eq:88}) and (\ref{eq:90}) into equation (\ref{eq:87}), we obtain
\begin{equation}
a\dot{H}=bH^{2}+cH+d, \label{eq:91}
\end{equation}
where we have defined
\begin{displaymath}
a\equiv1-\frac{3}{2}\xi_{2}, \qquad b\equiv\frac{3}{2}\left[3(\xi_{1}+\xi_{2})-(1+w)\right],
\end{displaymath}
\begin{equation}
c\equiv\frac{3}{2}\xi_{0} \quad \text{and} \quad d\equiv\frac{1}{2}(1+w)\Lambda-\frac{C}{12}(1-w) .\label{eq:92}
\end{equation}


Integration with $\tau(0) = 0$, $\tau(t_{0}) = 1$ and assuming $\xi_{1}+\xi_{2}<1/3$, gives 
\begin{equation}
H (t) =- \frac{c}{2b} + \hat{H}\coth{\left[-\frac{b}{a}\hat{H}t\right]}. \label{eq:94}
\end{equation}
where
\begin{displaymath}
	\hat{H}^{2} \equiv \left(\frac{c}{2b}\right)^{2}-\frac{d}{b}.
\end{displaymath}
The volume scale factor then takes the form
\begin{equation}
\tau(t) = \text{e}^{\frac{-3c(t-t_{0})}{2b}} \left(\frac{\sinh{\left[-\frac{b}{a}\hat{H}t\right]}}{\sinh{\left[-\frac{b}{a}\hat{H}t_{0}\right]}}\right)^{\frac{-3a}{b}},\label{eq:95}
\end{equation}
where
\begin{equation}
	t_{0} = -\frac{a}{b\hat{H}}\text{artanh}\left[\frac{\hat{H}}{H_{0}+\frac{c}{2b}}\right].\label{eq:96}
\end{equation}
In this case equations (\ref{eq:70}) and (\ref{eq:71}) reduce to 
\begin{equation}
A = \frac{C}{9H^{2}}, \label{eq:97}
\end{equation}
and 
\begin{equation}
	\rho = \rho_{0} + 3 (H^{2}-H_{0}^{2}), \label{eq:98}
\end{equation}
where $\rho_{0} = \rho(t_{0})$ is the energy density at the present time. If we assume that the bulk viscosity is constant, i.e. $\xi = \xi_{0}$, we see that for a Zel'dovich fluid with $w = 1$ equations (\ref{eq:94}) and (\ref{eq:95}) are identical to equations (\ref{eq:75}) and (\ref{eq:76}), respectively. In what follows we will look at the solutions for cold dark matter with the equation of state $p=0$.

\subsection{The case for pressureless matter, i.e. $w=0$.}

In this case equations (\ref{eq:94}) and (\ref{eq:95}) reduce to 
\begin{equation}
	H(t)=\frac{\xi_{0}}{2\left(1-3(\xi_{1}+\xi_{2})\right)} + \hat{H}\coth{\left[ \frac{3}{2}\frac{1-3(\xi_{1}+\xi_{2})}{1-\frac{3}{2}\xi_{2}}\hat{H}t\right]}, \label{eq:99}
\end{equation}
and
\begin{equation}
	\tau(t)=T \text{e}^{\frac{3\xi_{0}(t-t_{0})}{2\left(1-3(\xi_{1}+\xi_{2})\right)}}\sinh^{\frac{2(1-\frac{3}{2}\xi_{2})}{1-3(\xi_{1}+\xi_{2})}}{\left[\frac{3}{2}\frac{1-3(\xi_{1}+\xi_{2})}{1-\frac{3}{2}\xi_{2}}\hat{H}t\right]}, \label{eq:100}
\end{equation} 
where
\begin{equation}
T\equiv \left[\frac{H_{0}^{2}(3(\xi_{1}+\xi_{2})-1) + H_{0}^{2}\Omega_{\Lambda 0} + H_{0}\xi_{0} -\frac{C}{18} }{\frac{\xi_{0}^{2}}{4(3(\xi_{1}+\xi_{2})-1)}-H_{0}^{2}\Omega_{\Lambda 0}+\frac{C}{18}}\right]^{\frac{1-\frac{3}{2}\xi_{2}}{1-3(\xi_{1}+\xi_{2})}}. \label{eq:101}
\end{equation}
This expression generalizes the scale factor of the viscous isotropic FRW-universe model (see  ~\cite{nour}) to the volume scale factor of an anisotropic universe model.  In the limit that the bulk viscosity goes to zero, we obtain
\begin{equation}
	\tau(t)=\left[\frac{1-\Omega_{\Lambda 0}}{\Omega_{\Lambda 0}} \right] \sinh^{2}{\left(\frac{3}{2}\sqrt{\Omega_{\Lambda 0}}H_{0}t\right)}. \label{eq:102}
\end{equation} 
In the case of an isotropic universe this equation reduces to, see ~\cite{gron2002},
\begin{equation}
a(t)=\tau (t)^{1/3} =\left(\frac{\Omega_{m0}}{\Omega_{\Lambda 0}}\right)^{\frac{1}{3}}\sinh^{\frac{2}{3}}\left(\frac{3}{2}\sqrt{\Omega_{\Lambda 0}}H_{0}t\right),\label{eq:103}
\end{equation}
which is the expression for the scale factor for the $\Lambda  \textrm{CDM}$ model with no viscosity. Here $H_{0}$, $\Omega_{m0}$ and $\Omega_{\Lambda 0}$ are the values of the Hubble parameter, the density parameter of the dark matter and the density parameter of the dark energy at the present time, $t_{0}$, respectively.

The equation (\ref{eq:96}) for the age of the Universe in the presence of bulk viscosity reduces to
\begin{equation}
	t_{0} = \frac{2}{3\hat{H}}\frac{1-\frac{3}{2}\xi_{2}}{\left(1-3(\xi_{1}+\xi_{2})\right)}\text{artanh}\left[\frac{\hat{H}}{H_{0}+\frac{\xi_{0}}{2\left[3(\xi_{1}+\xi_{2})-1\right]}}\right].\label{eq:104}
\end{equation}
The age of the universe as a function of $\xi_{0}$ and $\xi_{1}$ is displayed in figure \ref{fig:fig10} with the value of $H_0$ determined from the condition that $t_{0} = 13.7\times 10^{9}$ years for $\xi = 0$. From this figure we see that when the viscosity increases, the age of the universe will also increase.

\begin{figure}[h!]
  \centering
  \subfigure[][$H(t)$.]{\label{fig:fig8a}\includegraphics[width=0.4\textwidth]{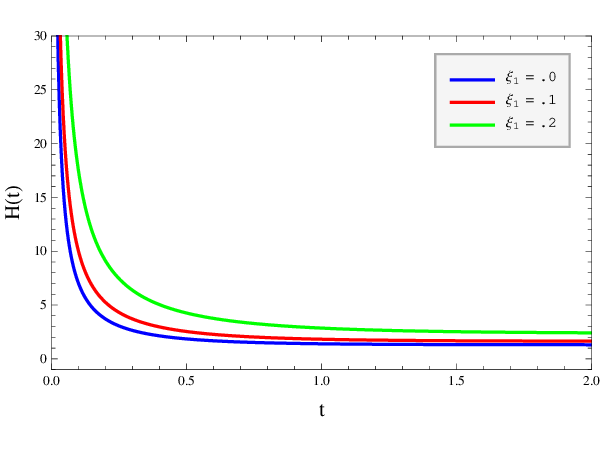}}
  \subfigure[][$\tau (t)$.]{\label{fig:fig8b}\includegraphics[width=0.4\textwidth]{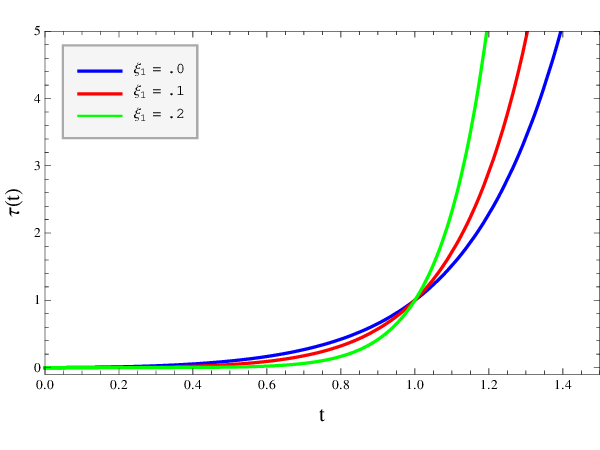}}
  \subfigure[][$\rho(t)$.]{\label{fig:fig8c}\includegraphics[width=0.4\textwidth]{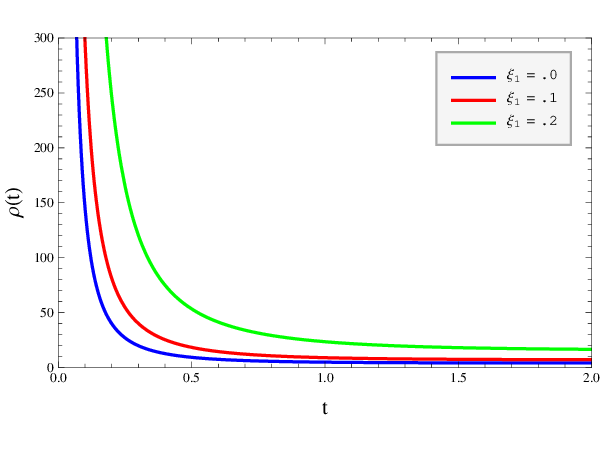}}
  \subfigure[][$A(t)$.]{\label{fig:fig8d}\includegraphics[width=0.4\textwidth]{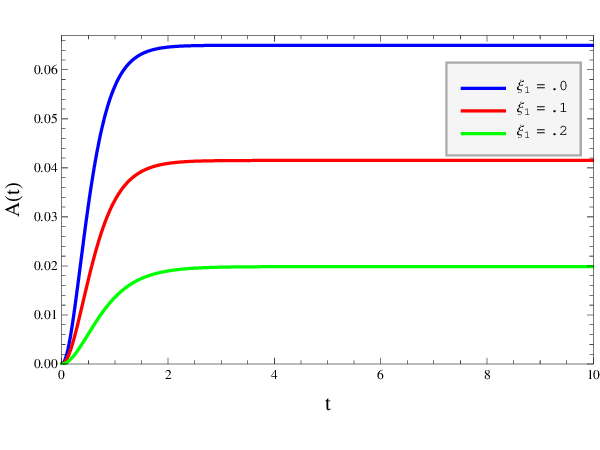}}
  \caption[scale factor]{The Hubble parameter, the scale factor, the energy density and the anisotropy parameter as functions of time. Here $\Omega_{\Lambda0} = 0.7$, $\xi_{2} = 0$, $\xi_{0} = 0.5$ and $C=H_{0} = t_{0} = 1$ for simplicity.}
  \label{fig:fig8}
\end{figure}

\begin{figure}[h!]
  \centering
  \subfigure[][$H(t)$.]{\label{fig:fig9a}\includegraphics[width=0.4\textwidth]{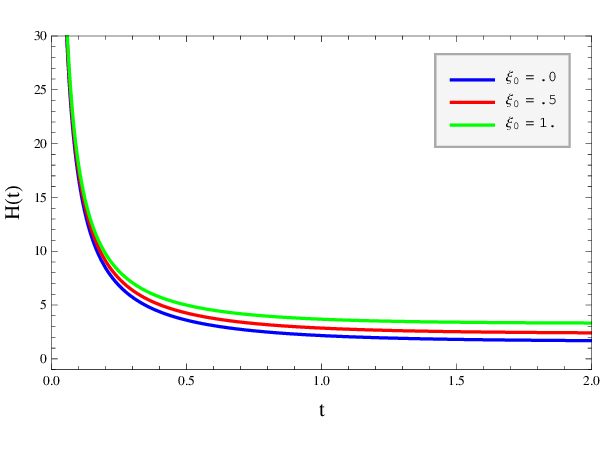}}
  \subfigure[][$\tau (t)$.]{\label{fig:fig9b}\includegraphics[width=0.4\textwidth]{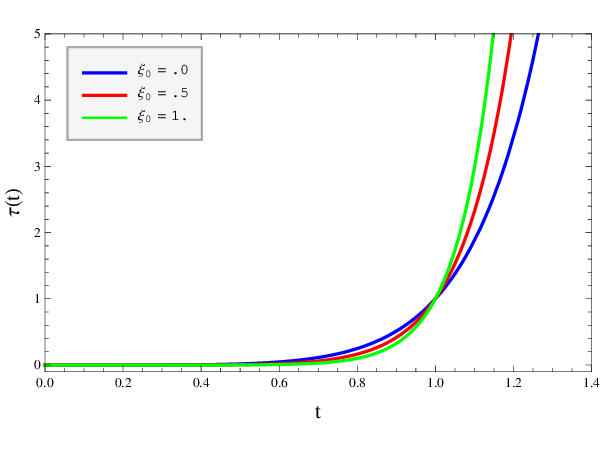}}
    \subfigure[][$\rho(t)$.]{\label{fig:fig9c}\includegraphics[width=0.4\textwidth]{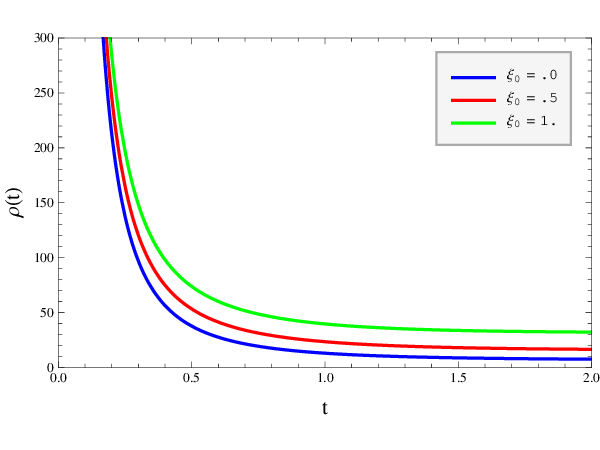}}
  \subfigure[][$A(t)$.]{\label{fig:fig9d}\includegraphics[width=0.4\textwidth]{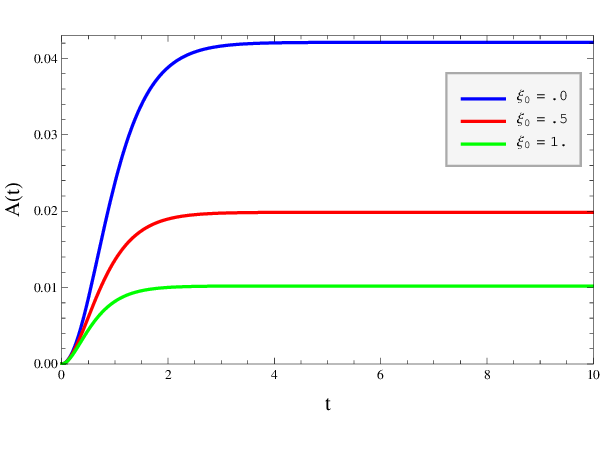}}
  \caption[scale factor]{The Hubble parameter, the scale factor, the energy density and the anisotropy parameter as functions of time. Here $\Omega_{\Lambda0} = 0.7$, $\xi_{2} = 0$, $\xi_{1} = 0.2$ and $C=H_{0} = t_{0} = 1$ for simplicity.}
  \label{fig:fig9}
\end{figure}

\begin{figure}[h!]
  \centering
   \includegraphics[width=0.45\textwidth]{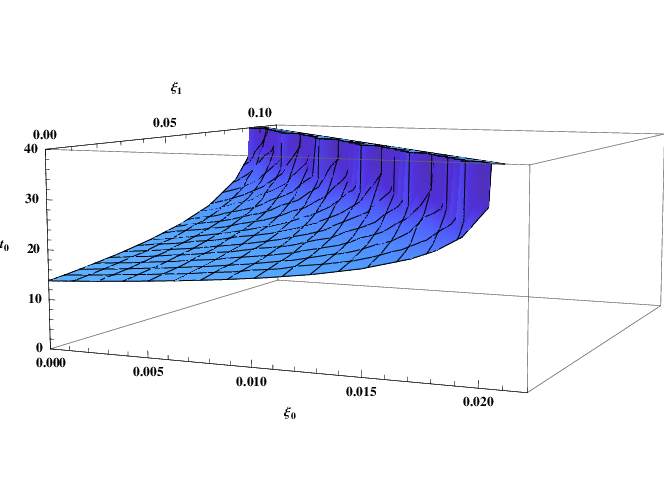}
  \caption[The age of universe]{The age of universe, $t_{0}$, as a function of bulk viscosity, i.e. $\xi_{0}$ and $\xi_{1}$. Here $\xi_{2}=0$.}
  \label{fig:fig10}
\end{figure}

In figures \ref{fig:fig8} and \ref{fig:fig9} we have assumed $\xi_{2}=0$ and we have plotted the Hubble parameter, the scale factor, the energy density and the anisotropy parameter for some different values of bulk viscosity, i.e. $\xi_{0}$ and $\xi_{1}$. From these figures we see that in this model the universe starts with a Big Bang at $t = 0$ with zero anisotropy. As $t$ increases the volume of this universe increases, but the energy density decreases. The energy density decreases faster for smaller values of bulk viscosity, which means that the bulk viscosity plays an important role in the energy production of the universe. The anisotropy parameter increases with time, but it will approach a finite value. The bigger the value of the bulk viscosity is the smaller is the anisotropy parameter. This means that bulk viscosity also contributes to isotropization of universe.

\begin{figure}[h!]
  \centering
    \subfigure[][The r-s-plane.]{\label{fig:fig10a}\includegraphics[width=0.45\textwidth]{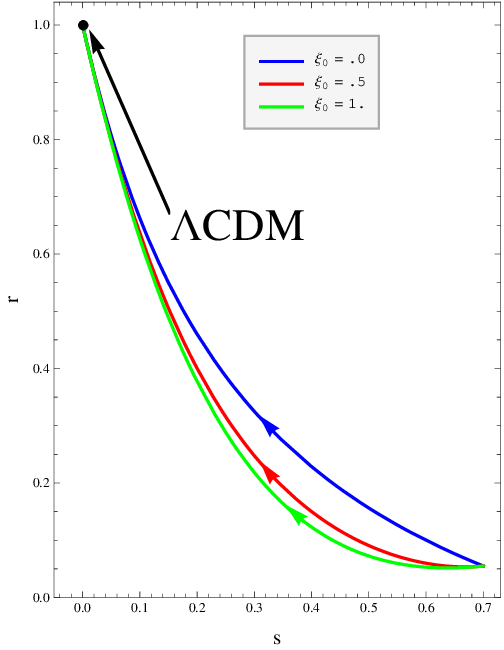}}
  \subfigure[][The r-s-plane.]{\label{fig:fig10b}\includegraphics[width=0.45\textwidth]{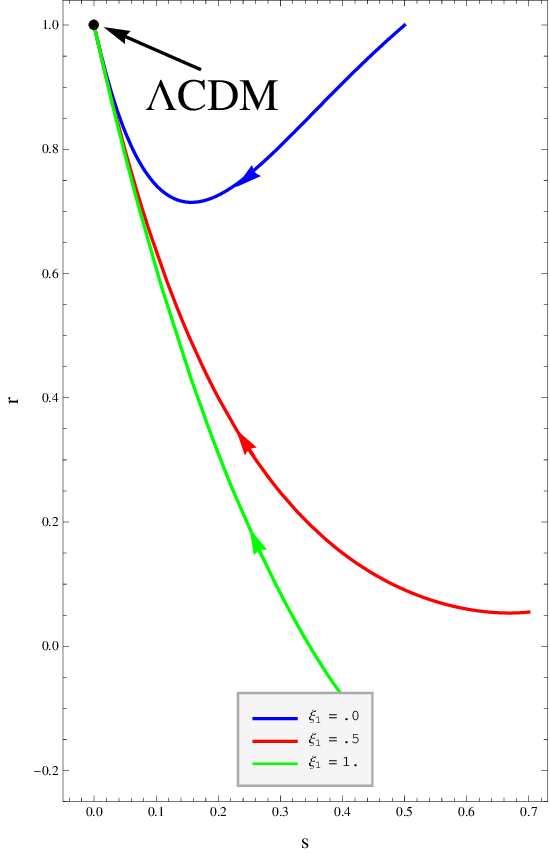}}
  \caption[The r-s-plane]{The r-s-plane. The point $\{0,1\}$ corresponds to $\{s,r\}$ for the $\Lambda\textrm{CDM}$ model with no viscosity.}
  \label{fig:fig11}
\end{figure}


\section{Results and Conclusion}

In this paper we have studied the Bianchi type-I universe models with nonlinear viscous fluid and applied the statefinder formalism to these models. By using a special relation between the bulk viscosity and the Hubble parameter and its derivatives, and by assuming that the shear viscosity is proportional to the Hubble parameter, we have given analytical solutions to the Raychaudhury and the continuity equation. The analytical solutions are given for Zel'dovich fluid with $w=1$ and for pressure-less matter with $w=0$. We have also shown that when the viscosity is set to zero we recover the standard $\Lambda$CDM model.

What we have found for these models is that the presence of bulk viscosity and the nonlinear viscous fluid will increase the energy density of matter. The evolution of the Hubble parameter, the volume scale factor and the anisotropy parameter will also depend on the bulk viscosity. If we increase the value of the bulk viscosity the anisotropy parameter will decrease.

From the results of this paper we can conclude that the Bianchi type-I universe models with nonlinear viscous fluid differ from the Bianchi type-I models with linear viscous fluid. This is because the nonlinear viscous fluid will increase the energy density of matter. From the plots of the r-s-plane we can also conclude that by use of statefinder parameter diagnostic method we can differentiate the Bianchi type-I universe models with nonlinear viscous fluid from other models.


\bibliographystyle{spr-mp-nameyear-cnd}

\end{document}